\documentclass[aps,twocolumn,superscriptaddress,prl,bibnotes,floatfix]{revtex4}
\usepackage{graphicx}

\def\apl{Appl.\ Phys.\ Lett.\ }

\def\prb{Phys.\ Rev.\ B }

\begin{document}

\title{AFM local oxidation nanopatterning of a high mobility shallow 2D
hole gas.}

\author{L.~P. Rokhinson}
\email[]{leonid@physics.purdue.edu}
\thanks{Present address: Department of Physics,
Purdue University, W. Lafayette, IN 7907 USA}
\affiliation{Department of Physics, Purdue University, West Lafayette, IN 47907 USA}
\affiliation{Department of Electrical Engineering, Princeton University, Princeton, NJ 08544 USA}

\author{D.~C. Tsui}
\affiliation{Department of Electrical Engineering, Princeton University, Princeton, NJ 08544 USA}

\author{L.~N. Pfeiffer}
\author{K.~W. West}
\affiliation{Bell Laboratories, Lucent Technologies, Murray Hill, New Jersey 07974 USA}

\begin{abstract}
Recently developed AFM local anodic oxidation (LAO) technique
offers a convenient way of patterning nanodevices, but imposes
even more stringent requirements on the underlying quantum well
structure. We developed a new very shallow quantum well design
which allows the depth and density of the 2D gas to be
independently controlled during the growth. A high mobility
($0.5\times 10^6$ cm$^2$/Vs at 4.2 K) 2D hole gas just 350 \AA\
below the surface is demonstrated. A quantum point contact,
fabricated by AFM LAO nanopatterning from this wafer, shows 9
quantum steps at 50 mK.
\end{abstract}

\maketitle

Rapidly developing area of physics of nanodevices laid down new
demands to heterostructure design, for example, shallow 2D gas is
required in order to improve size control and enhance the
confining potential. Even more stringent requirements are imposed
by recently developed AFM local anodic oxidation nanolithography
(LAO), which promise a low-energy high resolution alternative to
the conventional e-beam lithography\cite{sc,held97,bo02}. A key to
the success of LAO is the properly designed heterostructure. The
maximum depth of the LAO grown oxide is 100-150 \AA, and shifting
the surface by this amount should effectively deplete the
underlying 2D gas. Experimentally it has been deduced that 2D gas
should be $<400$ \AA\ below the surface\cite{held98}. In
conventionally grown shallow modulation doped quantum wells (QWs)
$\delta$-doping provides carriers to the 2D gas and compensates
the surface potential of GaAs.  There is a delicate balance
between underdoping, when most of the carriers are trapped at the
surface and no carriers left to form the 2D gas, and overdoping,
when $\delta$-doping layer becomes conducting at low temperatures.
An addition of another doping layer at the other side of the QW
reduces the necessary amount of the dopants in the top layer, but
precise control of the final density of the 2D gas is still
difficult.

We developed a novel heterostructure design which does not suffer
from either of the above mentioned drawbacks.  The as-grown
structure of a shallow p-type QW consists of a GaAs buffer grown
on a [$\bar{3}$11] substrate, followed by 3000 \AA\ AlGaAs
(x=0.32), Si $\delta$-doping ($6\times10^{11}$ cm$^{-2}$), 500
\AA\ AlGaAs setback, 150 \AA\ GaAs QW, 100 \AA\ AlGaAs barrier,
100 \AA\ GaAs, Si $\delta$-doping ($5\times10^{12}$ cm$^{-2}$),
and 150 \AA\ GaAs cap layer.  The calculated band diagram of this
structure is plotted in Fig.~\ref{ev} \cite{sim,tan90,snider90}.
The top $\delta$-doping resides in GaAs and serves the purpose of
compensating the surface potential and pinning the valence band
edge $E_V$ at the Fermi energy $E_F$. The doping should be low
enough to avoid conduction through this layer at low temperatures.
The design relaxes the criticality of the exact amount of doping
because this doping layer does not provide carriers to the
underlying 2D gas. The carriers are supplied by the bottom
$\delta$-doping. Again, the design provides some window for the
doping concentration because the density of the 2D gas is
determined by the setback thickness. An excessive doping leads to
the outdiffusion of dopants toward the QW and degradation of the
mobility\cite{pfeiffer91}.  Indeed, we have grown two wafers with
3 times different concentration of the bottom doping; both QWs
have hole concentrations different by $<5$\%, but the wafer with
lower doping has 8 time higher mobility.

\begin{figure}[t]
\def\ffile{ev}
\includegraphics[scale=0.9]{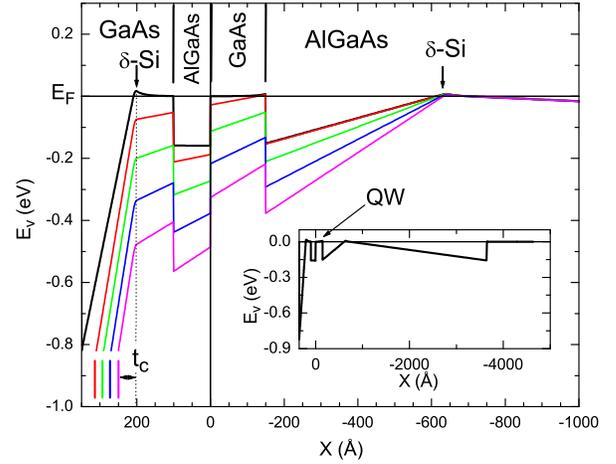}
\caption{Calculated band diagram for the heterostructure discussed
in the text.  Different curves correspond to the different
thicknesses of the cap layer $t_c=50$, 70, 90, 110 and 150 \AA;
vertical lines indicate position of the surface. The band diagram
for the full depth of the MBE growth, calculated for $t_c=150$
\AA, is shown in the inset.}
\label{\ffile}
\end{figure}

The band diagram is very sensitive to the thickness of the GaAs
cap layer $t_c$ and, thus, is ideal for the LAO nanolithography.
In Fig.~\ref{ev} we simulated the effect of the LAO by reducing
the $t_c$ by the oxide thickness. The 2D hole gas is depleted when
$t_c\sim110$ \AA, which corresponds to the oxide thickness of 60
\AA. 100 \AA\ of the oxide is estimated to lower the $E_V$ in the
QW region by 200 meV below the $E_F$, creating the corresponding
barrier for the adjacent 2D hole gas.

\begin{figure}[t]
\def\ffile{iv}
\includegraphics[scale=0.85]{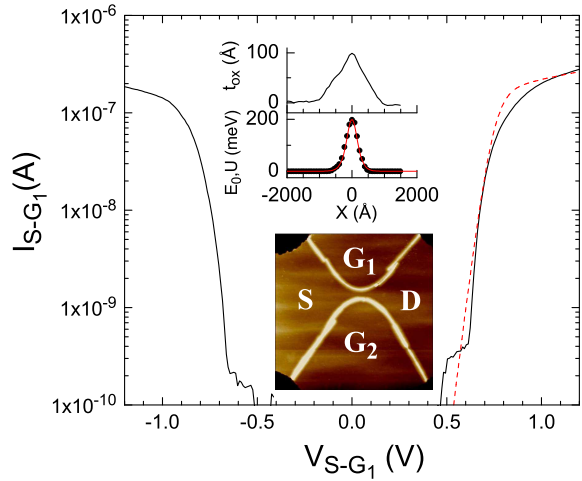}
\caption{Current-voltage (IV) characteristic across the oxide line
(solid line). Dashed line is the calculated IV using the model
potential discussed in the text. In the top inset AFM measured
oxide profile $t_{ox}$ and the corresponding calculated lowest
energy in the QW $E_0$ (dots), see Fig.~\ref{ev}, are plotted.
Solid line is the model potential $U(x)$ with $U_0=200$ meV, $V=0$
and $\alpha=45$ \AA$^{-1}$. The bottom inset is a $2 \mu$m~$\times
2 \mu$m AFM micrograph of the point contact, white lines are the
oxide lines, black corners are the edges of the mesa.}
\label{\ffile}
\end{figure}

We have grown a wafer using the above parameters. The wafer has
the hole density $1.38\times10^{11}$ cm$^{-2}$ and mobility
$0.48\times10^6$ cm$^{-2}$/Vs at 0.3 K, the highest reported for
so shallow (350 \AA\ below the surface) a 2D hole gas. An AFM
image of a quantum point contact, fabricated from this wafer using
AFM LAO, is shown in the inset in Fig. \ref{iv}. Profile of the
oxide is plotted in the top inset.  Position of the first energy
level $E_0$ in the QW, relative to the level position far from the
oxide line, is extracted from the band calculations and plotted
with black dots.  Due to non-monotonic dependence of the $E_0$ on
the oxide thickness, width of the potential barrier is narrower by
factor of 2 than the thickness of the oxide line and is
$\approx440$ \AA\ at half height. The current--voltage (IV)
characteristic across the oxide line, measured at 4.2 K, is shown
in Fig. \ref{iv}. Current is negligible at $|V|<0.6$ V and there
is a sharp turn on of the current at $\approx0.6$ V. We model the
IV characteristic by calculating the tunneling probability through
the barrier
\[U(x)=\frac{U_0}{\cosh^2(\alpha x)}+\frac{V/2}{\tanh(\alpha x)}, \]
where $U_0$ is the height of the potential under the center of the
oxide line, $V$ is the applied voltage, and $\alpha^{-1}$ is
related to the width of the oxide line. This potential\cite{llqm}
provides a good approximation to the expected potential, see solid
line in the inset in Fig. \ref{iv}. The turn on voltage is not
very sensitive to $\alpha$ at low temperatures for our barrier
widths, as well as to the exact shape of the potential. The
calculated current is plotted in Fig.~\ref{iv} with the dashed
line. The turn on voltage for $U_0=200$ meV is close to the
experimentally measure one. This value of $U_0$ is consistent with
the estimate of the tunneling barrier for 100 \AA\ oxidation from
band calculations. Experimentally, IV characteristic is linear for
oxidations depth $<60$ \AA, indicating that no tunneling barrier
is formed, also consistent with the band calculations.

Finally, we present transport characteristics of the point
contact. In Fig.~\ref{g-vg} differential conductance $G$ is
plotted as a function of the gate voltage $V_g=V_{g1}=V_{g2}$ at
50 mK. The $G$ is quantized in units of $2e^2/h$, and up to 9
plateaus can be clearly seen. Large number of steps indicate low
disorder in the vicinity of the point contact.  Thus, the designed
heterostructure, in combination with LAO technique, provides an
easy and reliable method of fabricating high quality
nanostructures with sub-nanometer control over the size, shape and
position.

\begin{figure}[ht]
\def\ffile{g-vg}
\includegraphics[scale=0.9]{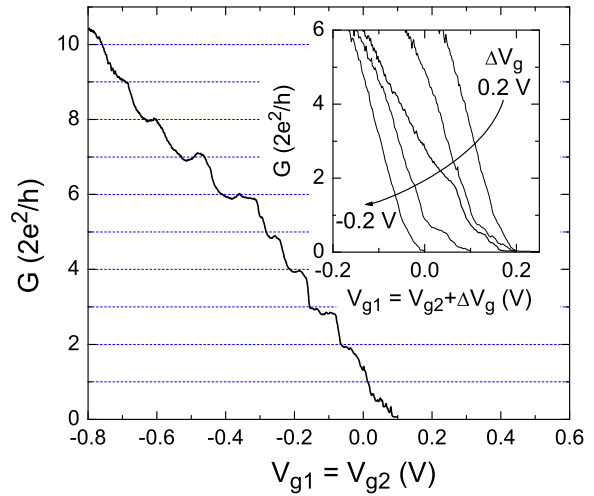}
\caption{Conductance $G$ as a function of the gate voltage
$V_g=V_{g1}=V_{g2}$ is measured at 50 mK.
In the inset $G$ is
measured at 4.2 K while a constant voltage difference $\Delta
V_g=-0.2$, -0.1, 0, 0.1 and 0.2, is applied between the gates. }
\label{\ffile}
\end{figure}

Previously, it has been shown that the position of the change can
be shifted by applying a voltage difference between the
gates\cite{heinzel00}.  Moreover, the barrier width under the
oxide and, correspondingly, the electrostatic coupling, can be
also adjusted.  The effect is the best demonstrated at elevated
temperatures, when steps are smeared out and the $dG/dV_g$ slope
reflects the electrostatic coupling between the gates and the 1D
channel.  In the inset in Fig.~\ref{g-vg} $G$ is plotted as a
function of $V_{g1}=V_{g2}+\Delta V_g$ for different values of
$\Delta V_g$ at 4 K . The slope $dG/dV_g$ increases with the
increase of $|\Delta V_g|$, indicating large electrostatic
coupling between one of the gates and the channel. This
enhancement is due to non-liner dependence of both channel and
barrier widths on the gate voltage for in-plane
gates\cite{larkin90}.

To summarize, we developed a novel heterostructure design for
reliable and reproducible growth of high mobility shallow QWs. We
demonstrate the design by fabricating a high quality quantum point
contact from such a wafer using AFM local anodic oxidation
technique. We show that amplitude of potential barriers, formed by
local oxidation of the surface, can be reliably predicted using
readily available numerical simulation tools.

We thank E.V. Tsiper for discussions. The work was supported by a
DURINT grant through the ONR and MARCO Focused Research Center on
Meterials, Structures, and Devices which is funded at the MIT, in
part by MARCO under contract 2001-MT-887 and DARPA under grant
MDA972-01-1-0035.


\end{document}